\documentclass[twocolumn,showpacs,prl,amsmath,amssymb,superscriptaddress]{revtex4}
\usepackage{graphicx}% Include figure files
\usepackage{bm}% bold math
\DeclareMathOperator{\Tr}{Tr} 

\DeclareMathOperator{\Max}{Max}

\begin{document}
%\draft 
%\tighten
%\onecolumn
%@twocolumnfalse\endcsname
\title{Quantum theory of spectral diffusion induced electron spin decoherence}
\author{W.M. Witzel} \affiliation{Condensed Matter Theory Center,
  Department of Physics, University of Maryland, College Park, MD
  20742-4111} \author{Rogerio de Sousa} \affiliation{Department of
  Chemistry and Pitzer Center for Theoretical Chemistry, University of
  California, Berkeley, CA 94720-1460} \author{S. Das Sarma}
\affiliation{Condensed Matter Theory Center, Department of Physics,
  University of Maryland, College Park, MD 20742-4111} \date{\today}
\begin{abstract}
A quantum cluster expansion method is developed for the problem of
localized electron spin decoherence due to dipolar fluctuations of lattice
nuclear spins.  At the lowest order it provides a microscopic
explanation for the Lorentzian diffusion of Hahn echoes without
resorting to any phenomenological Markovian assumption.  Our
numerical results show remarkable agreement with recent electron spin echo
experiments in phosphorus doped silicon.
\end{abstract}
\maketitle
%\vskip2pc]
%\narrowtext

%Intro
It was realized a long time ago that spectral diffusion due to the
dipolar fluctuations of nuclear spins often dominates the coherence
decay in electron spin echo experiments \cite{herzog56,klauder62}. The
recent advent of spin-based quantum computation in semiconductor
nanostructures revived the interest in spectral diffusion, which is
expected to be the dominant channel for low-temperature spin
decoherence in several spin-based quantum computer architectures
\cite{desousa03a}.  In spectral diffusion, the electron spin Zeeman
frequency diffuses in time due to the noise produced by the nuclear
spin bath.  Dipolar fluctuations in the nuclear spins give rise to a
temporally random effective magnetic field at the localized electron
spin, leading to 
%RdS: dephasing-> irreversible decoherence
irreversible decoherence
(i.e. a $T_2$-process).  All available
theories to date are based on classical stochastic modeling of the
nuclear field, a Markovian theoretical framework which is inevitably
phenomenological since it requires an arbitrary choice for the
spectrum of nuclear fluctuations.  Such a classical Markovian
modeling is arguably incompatible with the strict requirements of 
spin coherence and
control in a quantum information device. In addition, 
recent rapid experimental progress in
single spin measurements \cite{elzerman04}, which in the near future
promise sensitive measurements of quantum effects in spin resonance,
also warrant a quantum theory of spectral diffusion.
Here we propose a quantum theoretical framework for spectral diffusion
which is non-stochastic and fully microscopic.  In addition, our
theory produces an accurate quantitative prediction for the initial
decoherence, which is the most important regime for quantum
computation.  To the best of our knowledge, ours is the first quantum
theory for electron spin spectral diffusion.

Spectral diffusion is not a limiting decoherence process for silicon
or germanium based quantum computer proposals because these can, in
principle, be fabricated free of nuclear spins using isotopic
purification. Unfortunately this is not true for the important class
of materials based on III-V compounds, where spectral diffusion has
been shown to play a major role \cite{desousa03a,desousa03b}.  There
is as yet no experimental measurement of localized spin 
%RdS: dephasing-> decoherence (echo decay)
decoherence (echo decay) 
in
III-V materials, but such experimental results are anticipated in the
near future.

Our theory reveals that the inclusion of quantum corrections to
nuclear spin fluctuation increases the degree of decoherence,
%RdS: remove "at lower to intermediate time scales"
as is best evidenced from our explanation
of the existing factor of 
three discrepancy between the Markovian stochastic theory
\cite{desousa03b} and experimental data
\cite{chiba72,abe04,tyryshkinP} of spin echo decay in phosphorus doped
silicon.  Our method allows a fully microscopic explanation for the
observed time dependence of Hahn echo decay due to a nuclear spin
environment.  It was pointed out a long time ago
%RdS: removed chiba referencing from here because they did not fit
%exp-tau^2. 
\cite{klauder62,abe04} 
that the observed time dependence of
these echoes are well fitted to the expression $\exp{(-\tau^2)}$ 
(here $\tau$ is half of the time lag between the initial signal and an
echo), a behavior which can be derived phenomenologically by assuming
Lorentzian Brownian motion for the electron spin Zeeman frequency
%RdS: Added notefit here.
\cite{klauder62,notefit}.  In our method this behavior arises naturally from
the collective quantum evolution of the dipolar coupled nuclei,
without any phenomenological assumption on the dynamics of the
environment responsible for decoherence.  A proper description of
coupled spin dynamics is rather difficult due to the absence of Wick's
theorem for spin degrees of freedom. In that regard, variations of our
method may prove rather useful, since environmental spin baths are
ubiquitous in any device exploiting the coherent properties of quantum
spin systems.
 
%Formulation of the problem
The free evolution Hamiltonian for the spectral diffusion problem is
given by \cite{desousa03b}
\begin{equation}
{\cal H} = {\cal H}_{Ze} + {\cal H}_{Zn} + {\cal H}_{A} + {\cal H}_{B},
\label{H}
\end{equation}
where
\begin{eqnarray}
\label{H_Ze}
{\cal H}_{Ze} & = & \gamma_{S} B S_{z}, \\
\label{H_Zn}
{\cal H}_{Zn} & = & - \gamma_{I} B \sum_{n} I_{nz}, \\
\label{H_An}
{\cal H}_{A} & = & \sum_{n} A_{n} I_{nz} S_{z}, \\
\label{H_Bnm}
{\cal H}_{B} & = & \sum_{n \ne m} b_{nm}(I_{n+} I_{m-} - 2 I_{nz} I _{mz}).
\end{eqnarray}
Here $\bm{S}$ denotes the electron spin operator which couples to the
nuclear spin $\bm{I}_{n}$ located at the lattice site $\bm{R}_n$. The
nuclear spins are coupled to the electron through the hyperfine
constant $A_n$.  We have truncated Eq.~(\ref{H_An}) since the
non-secular hyperfine coupling can be neglected at moderate magnetic
fields ($B>0.1$~Tesla for the Si:P case). This interaction leads to
interesting effects at $B=0$ \cite{khaetskii02}, but at the moderate
magnetic fields required for spin resonance measurements it only
contributes a small visibility decay \cite{coish04}.  Each nuclear
spin is coupled to all others via the dipolar interaction
Eq.~(\ref{H_Bnm}), which is again truncated in the range of moderate
$B$ fields (For further details we refer to Ref.~\cite{desousa03b}).
The Hahn echo experiment consists in preparing the electron spin in
the initial state $|y_e\rangle=(\mid \uparrow\rangle + i \mid
\downarrow\rangle)/\sqrt{2}$, and then allowing free evolution for
time $\tau$. A $\pi$-pulse (here described by the Pauli operator
$\sigma_{x,e}$) is then applied to the electron spin, and after free
evolution for one more interval $\tau$ an echo is observed, which
provides a direct measurement of single spin coherence (i.e. $T_2$ or
$T_M$ in the usual 
%Rds: Typo: rotation->notation
notation).

We will now derive an exact expression for the Hahn echo decay due to
Eq.~(\ref{H}).  The density matrix (for electron and nuclear spins)
describing Hahn echo is given by
\begin{equation}
\rho(\tau) = U(\tau) \rho_{0} U^{\dag}(\tau),
\label{rho_t} 
\end{equation}
with the evolution operator 
\begin{equation}
U(\tau)=\textrm{e}^{-i {\cal H} \tau}\sigma_{x, e}\textrm{e}^{-i {\cal H} \tau}.
\end{equation}
Here $\rho_0$ is taken to be a thermal state for the nuclear spins, 
\begin{equation}
\rho_{0}=\frac{1}{2 M} |y_{e}\rangle\langle y_{e}| \otimes
\textrm{e}^{-{\cal H}_{n}/k_B T},
\label{rho_0}
\end{equation}
where ${\cal H}_n={\cal H}_{Zn}+{\cal H}_B$ and $M$ is its partition
function ($M\approx 2^N$ for $T\gg $~nK \cite{desousa03b}, where $N$
is the number of nuclear spins). The spin echo envelope is then given
by
%RdS: Wayne, please check a typo in this equation. In order to get 
%Eq 10 you need Sx-iSy in Eq. 9. This is minor and does not change the
%final result, but may be useful for people trying to derive it. 
%WMW: The absolute value signs make it correct either way.  Perhaps it
%is more confusing to derive in this form, but I'm not worrying about
%it now.
\begin{equation}
v_{E}(\tau) = 2 \left|\Tr{\left\{(S_x +i S_y)\rho(\tau)\right\}}\right|.
\label{amplitude_def}
\end{equation}
An explicit expression for Eq.~(\ref{amplitude_def}) can be obtained
by noting that the electron and nuclear spin Zeeman energies commute
with the total Hamiltonian, and $\sigma_{x,e}$ anticommutes with
$S_z$. After a few manipulations we get 
\begin{equation}
v_{E}(\tau)=\frac{1}{M}\left|\Tr{\left\{
U_{+}U_{-}\textrm{e}^{-{\cal H}_n/k_B T}
U_{+}^{\dag}U_{-}^{\dag}
\right\}} \right|,
\label{v_E}
\end{equation}
where 
\begin{equation}
U_{\pm}(\tau)=\textrm{e}^{-i{\cal H}_{\pm}\tau}
\label{Upm}
\end{equation}
are evolution operators under the effective Hamiltonians
\begin{equation}
{\cal H}_{\pm}={\cal H}_{B} \pm \frac{1}{2} \sum_{n} A_{n} I_{nz},
\label{hpm}
\end{equation}
which describe dipolar evolution under the effect of an electron spin
up (${\cal H}_{+}$) or down (${\cal H}_{-}$).  The trace in
Eq.~(\ref{v_E}) is taken over nuclear spin states only.

% Resubmission change
%added this paragraph
%Expansion in tau
In the high temperature limit ($k_B T \gg \gamma_{I} B$) we can expand
Eq.~(\ref{v_E}) in powers of $\tau$ to get
\begin{equation}
v_E(\tau) = 1 - \sum_{l=1}^{\infty} D_{2l}\;\tau^{2l}.
\label{tauexp}
\end{equation}
Defining the parameter 
\begin{equation}
c_{nm} = \frac{A_n - A_m}{4 b_{nm}},
\label{cnm}
\end{equation}
we obtain the first five coefficients $D_{2l}$ as a power
series of $c_{nm}$ and $b_{nm}$. For example, the first two
coefficients become explicitly $D_2=0$, $D_4=4 \sum_{n<m}c_{nm}^2
b_{nm}^4$.
%Wayne: I don't think there is need to show D_6 explicitly, takes too
%much space. Here it is (after correcting a typo)
%\begin{eqnarray}
%D_6 &=& -\frac{4}{3} \sum_{n<m} \left\{c_{nm}^4 b_{nm}^6\right.\\
%\nonumber
%&&+c_{nm}^2 b_{nm}^4 \left[b_{nm}^2 + \sum_{k\neq n,m}\left(b_{nk} - b_{mk}\right)^2\right]\\
%\nonumber
%&&\left.+\sum_{k\neq n,m}c_{nm} c_{nk} b_{nm}^2 b_{nk}^2
%\left[b_{nk} \left(b_{nm} + b_{mk}\right) + b_{mk}^2\right]\right\}.
%\end{eqnarray}
Truncating Eq.~(\ref{tauexp}) gives physical results only for
extremely short $\tau$ unless most nuclear pairs satisfy the condition
$c_{nm}\ll 1$. Nevertheless most physical problems are characterized
by several $c_{nm}\gg 1$, making evident the need for an alternative
expansion.  Thus the $\tau$ expansion, while being formally exact, is
not practical for coherence calculation except for extremely short $\tau$.

%Perturbation theory
In the $c_{nm} \gg 1$ regime, non-degenerate perturbation theory is
applicable to Eq.~(\ref{hpm}).  We introduce a bookkeeping parameter
$\lambda$ such that $\pm {\cal H}_{\pm}={\cal H}_0 \pm \lambda {\cal
  H}'$.  Here the unperturbed Hamiltonian ${\cal H}_0 =
\frac{1}{2}\sum_n A_n I_{nz}$ is diagonal in the nuclear spin $z$-basis,
while ${\cal H}'=\frac{1}{\lambda}{\cal H}_B$ is the dipolar
interaction rescaled to have the same magnitude as ${\cal H}_0$.
%WMW: more accurately stated to not equate A_n-A_m and A_n
A convenient choice for an order of magnitude estimate of 
$\lambda \sim 1/|c_{nm}|$ is to use the minimum possible value of 
$b_{nm}/|A_n-A_m|$ between nearest neighbors: 
$\lambda \sim Max(b_{nm})/Max(A_n) \sim 10^{-3}$ for Si:P.
Below we introduce a cluster expansion that can be related to powers
of $\lambda$ in this perturbation approach when $c_{nm}\gg 1$.
%

% Resubmission change
% 
% Change in nomenclature.  Call arbitrary, not necessarily connected,
% subsets a ``subset'' rather than a ``cluster''.  A ``cluster'' is
% now by definition (as it should be) connected.  Therefore you can
% only talk about ``clusters'' when we make the nearest neighbor
% approximation.  This makes the language simpler because instead of
% saying ``connected sub-clusters that are disconnected from each
% other'', you can simply say ``clusters contained in a subset''.

%Subset expansion
Let $\cal D$ be a subset of the nuclei in the problem.
Let $v_{\cal D}(\tau)$ be the solution of $v_{E}(\tau)$ 
%RdS:
[Eq.~(\ref{v_E})]
when only including the nuclei in $\cal D$.
We recursively define the contribution from subset ${\cal D}$
as $v_{\cal D}(\tau)$ minus contributions from any proper subset of 
${\cal D}$,
\begin{equation}
\label{subset_contrib}
v_{\cal D}'(\tau) = v_{\cal D}(\tau) - \sum_{{\cal S} \subset {\cal
    D}} v_{\cal S}'(\tau).
\end{equation}
For the empty set, we define $v_{\cal \emptyset}'(\tau) = v_{\cal
 \emptyset}(\tau) = 1$.
%RdS: Removed paragraph
Consider a subset contribution written
in the form of the non-degenerate perturbation expansion.  
Assuming
$\Max{(b_{nm})} \tau \ll
1$, we can show by the specific properties of ${\cal H}'$ that a 
cluster of size $k$ is composed of terms that are 
$O(\lambda^k)$ or higher.  In other words, we can write
the following expansion,
\begin{equation}
\label{subset_expansion}
v_E(\tau) = 1+\sum_{k = 2}^{k_0} \sum_{|{\cal D}| = k} v_{\cal D}'(\tau)
+ O(\lambda^{k_0+1}),
\end{equation}
where the second summation is over all possible nuclear subsets
of size $k$ (containing $k$ distinct nuclear sites).  
We note that a subset of size $1$ gives no contribution.
%

%Nearest neighbor
% Give some motivation by saying that \lambda N may not be small.
The $O(\lambda^{k_0+1})$ error in Eq.~(\ref{subset_expansion}) is
misleading because the number of terms of a given order of $\lambda$
may be large compared to $\lambda$.  The nature of this problem, as
well as a solution, becomes apparent when we use a nearest neighbor
approximation.  With this approximation, we ignore the interaction
between distant nuclei (i.e. pairs of nuclei for which $b_{nm}$ is
below some threshold) and divide our nuclear subsets into connected
``clusters.''  A subset contribution is then the product of its
cluster contributions.

% The argument here has changed a bit in this resubmission.
% Building a subset out of clusters of randomly chosen nuclei is a bit
% more logical.  It's still more lengthy than I'd like it to be, but
% hopefully it is a little easier to understand.
Consider all possible contributing subsets of size $k$.  We can categorize these
subsets by the number of clusters they contain.  To
estimate the number of subsets of size $k$ with $l$ clusters,
consider building the subset randomly.  First select $l$ nuclei at
random for each of the $l$ clusters.
The remaining $k-l$ nuclei are
chosen randomly from the neighbors of any previously chosen nuclei.  
Let $L$ be the average number of ``nearest neighbors'' for each nucleus.
The probability
that a nucleus will bridge two clusters that were meant to be separate
is at most $O\left(\frac{k L}{N}\right)$.  So the probability that any of the $k$ nuclei
will bridge two clusters is at most 
$O\left(1 - \left(1 - \frac{k L}{N}\right)^k\right)
\rightarrow O\left(\frac{k^2 L}{N}\right)$.  Therefore, as long as
$k^2 \ll N / L$, we can accurately estimate the number of subsets in
this manner of choosing nuclei at random.  Under this condition,
the number of subsets of size $k$ with $l$ clusters will scale roughly
(without dividing out permutations) as
$O(N^l L^{k-l})$, growing
exponentially with $N$ as we increase the number of clusters.
Clusters of size one give no contribution;
therefore, assuming $k \ll \sqrt{N/L}$ and $N \gg L$,
the possible contributing subsets of size $k$ are dominated by those containing all 
pairs (and a single triplet if $k$ is odd) which maximizes the number
of contained clusters. 
%RdS: Removed paragraph.
If $k = 2 l$ is even, then the number of contributing subsets of size
$k \ll \sqrt{N/L}$, largely composed of $l$ pairs, is $O\left(\frac{(N
  L)^{l}}{2^{l} l!}\right)$.  Therefore, our subset expansion [Eq.
(\ref{subset_expansion})] error grows with $k_0 = 2 l_0 \ll
\sqrt{N/L}$ as $O\left(\frac{(\lambda^2 N L / 2)^{l_0}}{l_0!}\right)$.
This is problematic because $\lambda^2 N L$ is not necessarily small.
Noting, however, that most contributing subsets are composed entirely
of pairs (except the one triplet of odd-sized subsets), we can
approximate the solution up to order
%RdS: << should be ~ to avoid confusion. When cluster size is
%~sqrt{N/L}, disconnected pairs do not dominate anymore.
$k_0 \sim \sqrt{N/L}$ 
by adding all possible products of pair
contributions as obtained by distributing the product
in the following,
% Resubmission change - wmw
% The error is actually O(N \lambda^3) because this is the lowest order
%
%RdS: No need to display \prod v_nm, equation looks ugly.
\begin{eqnarray}
\label{pair_prod}
v_E(\tau) &=& \prod_{n<m}\left\{1 + v_{nm}'(\tau)\left[1 + O(\lambda L)\right]\right\}\\ 
\nonumber
&&+ O\left(\frac{(\lambda^2 N L)^{l_0}}{2^{l_0} l_0!}\right),
\end{eqnarray}
which gives the lowest order of our cluster expansion.  
%RdS: Justification for neglecting the error above
The correction $O\left(\frac{(\lambda^2 N L / 2)^{l_0}}{l_0!}\right)$
is infinitesimal provided $l_0=k_0/2\sim \sqrt{N/L}\gg
\lambda^2N L$, or $\lambda\ll 1/(N L^3)^{1/4}$. For $N\sim 10^4$ and
$L\sim 10$, we get $\lambda\ll 0.02$ as the condition for disconnected
pairs to dominate spectral diffusion decay.
The $O(\lambda L)$ in Eq. (\ref{pair_prod}) represents the error
incurred by not considering clusters larger than pairs (including what
is required for odd subsets) and can be thought of as the contribution
you get by adding a neighbor to one of the distributed pairs.  Not all
of the terms obtained by distributing Eq.~(\ref{pair_prod}) will
contain pairs that are disconnected from each other.  However, using
the same argument we used to estimate numbers of contributing subsets,
$l < k_0/2$ random pairs will most likely be disconnected when
%RdS: Again changed << to ~
$k_0\sim \sqrt{N/L}$.  
These extraneous terms are therefore negligible at
each order below $k_0$.
%RdS: Moved exponential approximation below, after exact solution is
%presented. This makes its justification more straightforward.
%
% This text was removed -- Justification already given above. 
%
%which is as valid as Eq. (\ref{pair_prod}) with our $k_0 \ll
%\sqrt{N/L}$ requirement.  In this form, it is easier to see the effect
%of the error imposed by a finite $k_0 = 2 l_0$.  By our initial
%conditions, $v_2(0) = 0$.  As $\tau$ increases, $v_2(\tau)$ will
%approach something on the order of $-\lambda^2 N L$.  In order for
%Eq.~(\ref{exp_pairs}) to exhibit a full decay, we must have $\lambda^2
%N L \gg 1$.  If $k_0 \gg 1$, then the right side of Eq.
%(\ref{exp_pair_series}) gives a series that converges well as long as
%$v_2(\tau) \ll k_0$.  This implies our approximate solution is good
%down to the tail of the decay, approaching $e^{-k_0}$ or
%$e^{-\sqrt{N/L}}$.  Since we assume $N \gg L$ in our use of the
%nearest neighbor approximation, this tail error is insignificant.

For a cluster of two nuclear spins (${\cal D}=\{n,m\}$) exact
evaluation of $v_{\cal D}(\tau)$ using Eq.~(\ref{v_E}) in the high
temperature limit leads to
\begin{eqnarray}
v_{nm}(\tau) &=& 1+ v_{nm}'(\tau)\nonumber\\
&=&1 - \frac{c_{nm}^2}{(1 + c_{nm}^2)^2}
\left[\cos{(\omega_{nm} \tau)} - 1\right]^2, \label{vnm}\\
\omega_{nm} &=& 2 b_{nm} \sqrt{1 + c_{nm}^2},
\end{eqnarray}
with $c_{nm}$ defined in Eq. (\ref{cnm}).
%RdS: Exponential approximation
Using Eq.~(\ref{pair_prod}) and the condition $\Max{(b_{nm})}\tau\ll 1$
we write the final expression for the lowest order cluster expansion as
%RdS: No need to define v_2 anymore
\begin{equation}
\label{exp_pairs}
v_E(\tau) \approx \exp{\left\{\sum_{n < m} v_{nm}'(\tau)\left[1 + O(\lambda L)\right]\right\}}.
\end{equation}
%RdS: Added comment on lambda L error.
Note that including clusters of three adds a correction $O(\lambda L)$
to the decay.
%

% Resubmission change
% addition
%RdS: Wayne, I adopted your suggestions. 
We have presented two theories.  Both require $\tau \ll
\Max{(b_{nm})}^{-1}$; however, in problems we've considered, the decay time
is well within this limit. We argued the $\tau$~expansion
[Eq.~(\ref{tauexp})] converges for $c_{nm} \ll 1$ while the cluster
expansion [Eq.~(\ref{pair_prod})] for $c_{nm} \gg 1$. The cluster
expansion becomes non-perturbative through the use of the exact
solution Eq.~(\ref{vnm}).  In fact, in the same way that cluster size
was related to minimum orders of $\lambda$ in the perturbation
expansion, we can also relate cluster size to minimum orders of $\tau$
in the $\tau$-expansion.  For example, by taking $c_{nm}\ll 1$ in
Eq.~(\ref{vnm}) we recover Eq.~(\ref{tauexp}) to lowest order, showing
that this exact solution interpolates between the two regimes at
lowest order.  For physical problems where a wide range of parameters
$c_{nm}$ coexist, exact evaluation of larger clusters provides a novel
systematic approximation to the problem of spectral diffusion.

\begin{figure}
\includegraphics[width=3in]{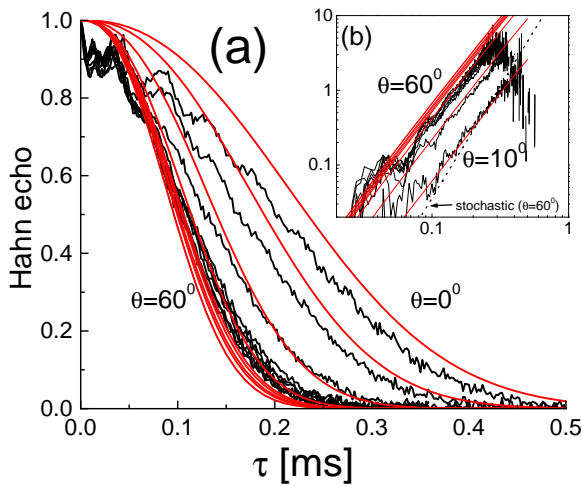}
\caption{
  Hahn echo decay $v_E(\tau,\theta)$ of a phosphorus donor electron
  spin in silicon. (a) Theory (solid lines) and experiment is shown
  for several orientation angles of the magnetic field with respect to
  the crystal lattice, ranging from the [100] to the [110] direction
  ($\theta=0,10,20,\ldots,90$).  (b) Here we plot
  $-\ln{v_E}(\tau,\theta)+\ln{v_E(\tau,\theta=0)}$, allowing for the
  removal of any decoherence mechanism which is independent of
  $\theta$. The qualitative and quantitative agreement between theory
  and experiment is remarkable, in contrast to the stochastic approach
  (dashed).
\label{fig1}}
\end{figure}

We use Eqs. (\ref{exp_pairs}) and (\ref{vnm}) to perform explicit
calculations of electron spin echo decay of phosphorus impurities in
silicon \cite{chiba72,abe04,desousa03b}. In this case the parameter
$A_n$ is given by the hyperfine shift of a nuclear spin positioned a
vector $\bm{R}_n$ from the donor center,
%RdS: Corrected typo, psi->Psi
\begin{equation} 
A_{n}=\frac{8\pi}{3}\gamma_{S}\gamma_{I}\hbar|\Psi(\bm{R}_{n})|^{2}.
\end{equation}
We used $\gamma_{S} = 1.76 \times 10^7 \mbox{(s G)$^{-1}$}$ and
$\gamma_{I} = 5.31 \times 10^3 \mbox{(s G)$^{-1}$}$.  Here $\Psi(\bm{R}_n)$
is the Kohn-Luttinger wave function of a phosphorus donor impurity in
silicon, 
%RdS: removed ``further details''
as described in Ref.~\cite{desousa03b}. The
central $^{31}$P nuclear spin does not contribute to spectral
diffusion because its hyperfine energy is significantly larger than
any of its neighbors, suppressing the spin flips by energy
conservation.  Dipolar coupling is given by
\begin{equation}
b_{nm}=-\frac{1}{4}\gamma_{I}^{2}\hbar\frac{1 - 3 \cos^{2}{\theta_{nm}}}{R^{3}_{nm}}.
\end{equation}
It contains an important anisotropy with respect to the angle
$\theta_{nm}$ formed between the applied magnetic field and the bond
vector linking the two spins ($\bm{R}_{nm}$). This property leads to a
strong dependence of spin echo decay when the sample is rotated with
respect to the applied B field direction. Fig.~\ref{fig1} shows
experimental data when the sample is rotated from the [100] to the
[110] direction. Here the cluster approximation is expected to be
appropriate for $\tau\ll 1-5$~ms. Finally, in a natural sample of
silicon only a small fraction $f=4.67$\% of lattice sites 
have non-zero nuclear spin (these are the spin-$1/2$ $^{29}$Si isotopes).
Averaging Eq.~(\ref{pair_prod}) we get 
\begin{equation}
v_{E}(\tau)=\prod_{n < m}\left[
(1-f^2)+f^2 v_{nm}(\tau)\right].
\label{vef}
\end{equation}

Our numerical calculations of Hahn echo decay for several magnetic
field orientation angles are shown on Fig.~\ref{fig1}(a). We also show
experimental data taken for bulk natural silicon with phosphorus
doping concentration equal to $2\times 10^{15}$~cm$^{3}$
\cite{tyryshkinP}.  The high concentration of phosphorus donors leads
to an additional decoherence channel arising from the direct spin-spin
coupling between the electron spins that contribute to the echo.  This
contribution can be shown to add a multiplicative factor
$\exp{(-\tau/1\,\rm{ms})}$ to Eq.~(\ref{vef}) \cite{abragam62}.
Because this contribution is independent of the orientation angle, we
can factor it out by subtracting the $\theta=0$ contribution from the
logarithm of the experimental data taken at angle $\theta$. The result
is shown on Fig.~\ref{fig1}(b) (log-log scale). Our theory seems to
explain the time dependence of the echo quite well.  To check
convergence of our cluster expansion we have gone to the next order.
Including clusters of three amounts to a contribution of $0.1\%$,
in agreement with our estimate of $\lambda f L \sim 10^{-3}$.  
We have also verified that our cluster expansion results agree
quantitatively with Eq.~(\ref{tauexp}) for small $\tau$ when excluding
nuclei close to the center of the electron wave function where $c_{nm}\gg1$.  This result
is to be compared with the recent stochastic theory developed by two
of us \cite{desousa03b} (Dashed line in Fig.~\ref{fig1}(b) shows the
stochastic calculation for $\theta=60^{\circ}$). Although the
stochastic theory yields correct order of magnitude for the coherence
times, it fails qualitatively in explaining the time dependence.  The
present method is able to incorporate all these features within a
fully microscopic framework.

An important issue in the context of quantum information is the
behavior of spin coherence at the shortest time scales. The
experimental data in Fig.~\ref{fig1} reveals several oscillating
features which are not explained by our current method. These are echo
modulations arising from the anisotropic hyperfine coupling omitted in
Eq.~(\ref{H}) \cite{abe04}.  This effect can be substantially reduced
by going to higher magnetic fields (In a quantum computer
$B\sim9$~Tesla will probably be required in order to avoid loss of
fidelity due to echo modulation \cite{saikin03}). On the other hand,
spectral diffusion is essentially independent of magnetic field even
to extremely high values ($B\sim 10$~Tesla).  Nevertheless this effect
%RdS: Added ``is''
is 
expected to be absent in III-V materials \cite{yablonovitch03}, hence
our theory allows the study of spin coherence at time scales of great
importance for quantum information purposes but currently inaccessible
experimentally.

%Conclusion
In conclusion, we describe a new quantum approach for the problem of localized
electron spin decoherence due to the fluctuation of dipolar
coupled nuclear spins. In contrast to former theories, our method
requires no {\it ad hoc} stochastic assumption on the complex dynamics
of the environment responsible for decoherence. Hence it provides an
important example where direct integration of the environmental
equations of motion provides a systematic understanding of the loss of
coherence which needs to be controlled for quantum information
applications.   We are indebted to A.M. Tyryshkin and S.A. Lyon for providing the
experimental data shown in Fig. \ref{fig1}.  This work is supported by
ARDA, ARO, and NSA-LPS.

After the completion of the review process of our manuscript, a
preprint by Wang Yao, Ren-Bao Liu, and L. J. Sham \cite{yao} appeared
exactly reproducing our lowest order theoretical result by a
completely different technique which treats excitations of
nuclear pair-correlations as quasi-particles that are non-interacting 
for $\tau \ll \Max{(b_{nm})}^{-1}$.  This independent agreement demonstrates the
validity of our cluster expansion technique.

%
%%%%%%%%%%%%%%%%%%%%%%%%%%%%%%%%%%%%%%%%%%%%%%%%%%%%%%%%%%%%%%%
%


\begin{thebibliography}{99}

\bibitem{herzog56} B. Herzog and E.L. Hahn, Phys. Rev. {\bf 103}, 148
(1956).

\bibitem{klauder62} J.R. Klauder and P.W. Anderson, Phys. Rev. {\bf
  125}, 912 (1962).

\bibitem{desousa03a} R. de Sousa and S. Das Sarma, \prb {\bf 67},
033301 (2003).

\bibitem{elzerman04} J.M. Elzerman {\it et al.}, Nature {\bf 430}, 431
(2004); M. Xiao {\it et al.}, Nature {\bf 430}, 435 (2004); D. Rugar
{\it et al.}, Nature {\bf 430}, 329 (2004).

\bibitem{desousa03b} R. de Sousa and S. Das Sarma, \prb {\bf 68},
115322 (2003).

\bibitem{chiba72} M. Chiba and A. Hirai, J. Phys. Soc. Japan {\bf 33},
730 (1972).

\bibitem{abe04} E. Abe, K.M. Itoh, J. Isoya and S. Yamasaki, \prb {\bf
  70}, 033204 (2004).

\bibitem{tyryshkinP} A.M. Tyryshkin and S.A. Lyon, Private
communication; 
A.M. Tyryshkin, S.A. Lyon, A.V. Astashkin, A.M. Raitsimring, \prb {\bf 68},
193207 (2003).

%RdS:
\bibitem{notefit} Refs. \cite{chiba72,tyryshkinP} used a two parameter
fit $v_E=\exp{(-\tau-\tau^3)}$, which is strongly affected by the echo
modulations. As noted in Ref. \cite{abe04} and in our Fig. 1(b)
$\exp{-\tau^2}$ provides a better fit.

\bibitem{khaetskii02} A.V. Khaetskii, D. Loss, L. Glazman \prl {\bf
  88}, 186802 (2002). 

\bibitem{coish04} N. Shenvi, R. de Sousa, and K. B. Whaley, \prb
{\bf 71}, 224411 (2005).

\bibitem{abragam62} A. Abragam, {\it The Principles of Nuclear
Magnetism} (Oxford  University Press, London, 1961), chapter IV,
Eq.~(63).

\bibitem{saikin03} S. Saikin, L. Fedichkin, \prb {\bf 67}, 161302(R) (2003).

\bibitem{yablonovitch03} E. Yablonovitch {\it et al.}, Proc. of the
IEEE, {\bf 91}, 761 (2003).

\bibitem{yao}Wang Yao, Ren-Bao Liu, and L. J. Sham, cond-mat/0508441
  (2004).

\end{thebibliography}
\end{document}